\documentstyle[aps,pre,multicol,epsf]{revtex}
\def\be{\begin{equation}}
\def\ee{\end{equation}}

\def\bea{\begin{eqnarray}}
\def\eea{\end{eqnarray}}

\begin{document}

\title{Derivation of the Matalon-Packter law
for Liesegang patterns}
\vspace {1truecm}

\author{T. Antal${}^1$, M. Droz${}^2$, J. Magnin${}^2$, 
Z. R\'acz${}^{1,3}$, and M. Zrinyi${}^4$}
\address{{}$^1$Institute for Theoretical Physics,
E\"otv\"os University,
1088 Budapest, Puskin u. 5-7, Hungary}
\address{{}$^2$ {D\'epartement de Physique Th\'eorique, Universit\'e de 
Gen\`eve,
CH 1211 Gen\`eve 4, Switzerland.}}
\address{${}^3$Department of Physics, Theoretical Physics, Oxford University\\
1 Keble Road, Oxford, OX1 3NP, United Kingdom}
\address{${}^4$Department of Physical Chemistry, Technical University of 
Budapest, H-1521 Budapest, Hungary}
\date{Preprint: \today}
\maketitle
\begin{abstract}
Theoretical models of the Liesegang phenomena are studied and 
simple expressions for the spacing coefficients characterizing the patterns
are derived. The emphasis is on displaying the explicite dependences on 
the concentrations of the inner- and the outer-electrolytes. Competing 
theories (ion-product supersaturation, nucleation and droplet growth, 
induced sol-coagulation) are treated with the
aim of finding the distinguishing features of the theories. 
The predictions are compared with experiments and the results 
suggest that the induced sol-coagulation theory is the best candidate
for describing the experimental observations embodied 
in the Matalon-Packter law.\\
\begin{center}
\bf{UGVA DPT 1998/07-1012}
\end{center}
\end{abstract}
\pacs{}

\date{\today}

\maketitle
\begin{multicols}{2}
\narrowtext
\section{Introduction}

Systems that exhibit pattern formation are common in nature and 
patterns often emerge in the wake of a moving front \cite{Hoh}. 
A well known example that will be studied below
is the Liesegang phenomenon~\cite{{liese},{Henisch}}, the formation of  
precipitation patterns by a moving chemical reaction front. 
In a typical experimental setup, one has a
chemical reactant (inner electrolyte) dissolved in a gel matrix and
a second reactant (outer electrolyte) poured onto the gel. 
The concentration
of the outer electrolyte is much higher than that of the inner one
and so it diffuses into the gel
and reacts with the inner reactant. The reaction product then
precipitates and, frequently, the observed precipitation 
patterns are a family of
bands or rings (depending on the geometry of the
system) clearly separated in the direction perpendicular to the
diffusion. This process is believed to be responsible for many 
precipitation patterns such as 
the structure of agate rocks \cite{Henisch} or, to cite 
a somewhat speculative example, the pattern of human gallstones. 

Although the Liesegang patterns have been known for more than one hundred 
years~\cite{liese},  there is still disagreement as to the mechanisms 
underlying this phenomenon. The main problem 
is the lack of a theory that could be compared with 
experiments quantitatively and, to a lesser degree, the lack of 
experiments that produce results amenable to  
quantitative theoretical analysis. At first the task appears to be simple. 
The Liesegang bands are formed at times $t_n$ at some positions $x_n$ 
(measured from the planes separating the reagents at the initial moment),
and have a width $w_n$. The quantities $t_n\, ,\, x_n$ and $w_n$
obey the following generic laws:

i) After a transient time,  the position $x_n$ of the $n$-th 
band is related to
the time $t_n$ of its formation through the so-called {\it time law}, $x_n\sim 
\sqrt{t_n}$, first discussed in \cite{morse}. The time law is a rather 
obvious consequence of the diffusive motion of a reaction front 
in a gel and thus it is considered to be understood. 

ii) For large $n$, the ratio of the positions of two
consecutive bands approaches a constant value 
\be
{{x_n}\over{x_{n-1}}} \to 1+p
\label{1+p}
\ee 
with $0.05<p<0.4$ typically. 
This last property is known as the Jablczynski law~\cite{jabli} or the {\it 
spacing law} and $1+p$ is called the {\it spacing coefficient}.
Most of the detailed experimental observations concern this law. It 
has been found that the spacing coefficient is a nonuniversal quantity. 
It depends for example on the concentrations $a_0$ and $b_0$ of the outer and 
inner electrolytes which can be controlled experimentally.
Based on several different experiments, Matalon and 
Packter~\cite{{Matalon},{Packter}} concluded that:
\begin{equation}
p=F(b_0) + \frac{G(b_0)}{a_0} \quad ,
\label{MatPac}
\end{equation}
where $F$ and $G$ are decreasing functions of their 
arguments. For various reagents, they found $F(b_0)\sim b_0^{-\gamma}$  
with $0.2\le \gamma\le 2.7$. The function $G(b_0)$ is less known but it
is generally observed that it decreases with increasing $b_0$. 
The observation expressed in equation 
(\ref{MatPac}) is usually called the {\it Matalon-Packter law}.

A problem with the Matalon-Packter law is that, experimentally, 
$a_0$ and $b_0$ can usually be changed only by a factor 5-10 and, 
furthermore, $p$ itself has a significant error bar since it is 
determined from a finite number of $x_n$-s ($n\sim 10-20$). Thus the
functional form (\ref{MatPac}) is far from unique, it should be considered
as a power law fit to $p$ \cite{revertnote}.

iii) Finally, the {\it width law} states that the width $w_n$ of 
the $n$-th band is an increasing function of $n$ and 
typically $w_n \sim x_n^{\alpha}, ~\alpha > 0$ \cite{widthlaw}. Since 
the definition of $w_n$ is open to debate 
and the error bars in the measurements of $w_n$ are rather large, 
this quantity has largely been ignored in the quantitative discussions 
of Liesegang phenomena. 

Although there are a large number of interesting and often puzzling 
observations about Liesegang phenomena in particular systems, 
the above three points appear to be the only ones which 
describe quantifiable results and carry some generality. 
 
One would expect that so few experimental facts can be explained easily.
Indeed, there are several competing theories 
\cite{{Ostwald},{Wagner},{Prager},{zeldo},{Chatter},{shino},{dee},{luthi},{flicker},{kaimul},{ortoleva},{venzl}} many of which 
\cite{{Ostwald},{Wagner},{Prager},{zeldo},{Chatter},{shino},{dee},{luthi}}
fare well as far
as the derivation of the time and spacing laws are concerned \cite{postnuc}. 
All these theories follow how the 
diffusive reagents (ions in the outer and inner electrolytes) 
$A$ and $B$ turn into immobile precipitate $D$
\be
A+B\rightarrow ... C ... \rightarrow D
\label{process}
\ee
by taking into account various scenarios for the intermediate steps 
denoted as $...C...\,$. Since the precipitate appears 
through some kind of supersaturation, further differences in theories
arise from the details of treating the nucleation
thresholds and the growth kinetics  of the precipitate.

The simplest (and first developed) theory is 
based on the concept of {\it supersaturation of ion-product}~\cite{Ostwald}
and has been developed by many researchers
\cite{{Wagner},{Prager},{zeldo},{Smith}}. In this theory, there is 
no intermediate step in-between $A,B$ and $D$. 
When the local product of the concentrations of the reactants, $ab$, 
reaches some critical value, $q^*$, nucleation of the precipitate $D$ occurs. 
The nucleated particles grow and deplete $A$ and $B$ in their 
surroundings. As a consequence, 
the local level of $ab$ drops and no new nucleation takes place. 
This continues until the reaction zone (where $ab$ is maximum) 
moves far enough that the depletion effect of the 
precipitate becomes weak. Then $ab=q^*$ is reached again 
and nucleation can occur.
The repetition of this process leads to the formation of bands.

In the next level of complexity, theories
contain a single intermediate step in $...C...$ 
with the mechanism of band
formation based on the supersaturation of the intermediate 
compound $C$~\cite{dee,luthi}.   
It is assumed that $A$ and $B$ react to
produce a new species $C$ which also diffuses in the gel. 
The nature of $C$ is not really specified, 
it may be a molecule as well as a colloid particle. When the
local concentration of $C$ reaches some threshold value, nucleation
occurs and the nucleated particles, $D$, act as aggregation seeds.
The $C$ particles near to $D$ aggregate to the existing droplet 
(hence become $D$) 
provided their local concentration is larger 
than a given aggregation threshold. 
These models are characterized by two thresholds, 
one for nucleation and one for droplet growth. 
The depletion mechanism is similar to the one 
described for the ion-product theory and it leads to band formation.
We shall refer to this theory as the theory 
of {\it nucleation and droplet growth}.

A variation of the single intermediate compound theories is 
based on the idea of an {\it induced sol-coagulation}
process \cite{{Chatter},{shino}}. 
The compound $C$ is assumed to be the sol and this sol coagulates if the
following two conditions are satisfied.
First, the concentration of $C$  exceeds a supersaturation 
threshold $c\ge c^*$ and, second,
the local concentration of the outer electrolyte is above a threshold
$a>a^*$ that is often referred to as the critical coagulation concentration 
threshold.
The band formation is a consequence of the nucleation 
and growth of the precipitate combined with the motion of the front 
where $a=a^*$.

The time- and the spacing laws follow from the above theories. Thus, 
to select the correct theory (provided there is a single theory
for all Liesegang phenomena), one would have to 
calculate the functional form of $p(a_0,b_0)$ 
in order to find agreement or disagreement with the Matalon-Packter law.
The spacing coefficient, however, is obtained from the numerical 
solution of complicated coupled nonlinear differential 
equations and, as a consequence, 
the results are in an implicit form that is not 
particularly useful for deducing $p(a_0,b_0)$. A notable exception
is the simplest version of the ion-product theory which was used by 
Wagner \cite{Wagner} to derive a result, $p\sim a_0^{-2/3}$, that is at 
variance with the Matalon-Packter law. 
Wagner's result raises the question whether the present theories
contain the Matalon-Packter law at all and our aim here is to discuss
this question and make an attempt at answering it.

In attempting  to derive the Matalon-Packter law, our basic aim is to
keep the calculations and formulas simple so that the
explicit dependence on $a_0$ and $b_0$ could be displayed. 
Accordingly, we have to make simplifying assumptions. 
The first two of these assumptions 
are about the experimental setup and 
the majority of the experiments satisfy the constraints described there.
The remaining assumptions 
are based on observations of the time evolution of the 
concentrations of the reactants and reaction products 
obtained from numerical solution of the 
appropriate equations, from simulations of the process 
\cite{luthi}, and also from experience with the analytical solution 
of the problem of chemical reaction zone \cite{GR}.
We believe that the important features of the original problem 
are not lost after making the assumptions listed below. 

\begin{enumerate}

\item The experimental setup can be described by one-dimensional 
reaction-diffusion equations (the gel is in a testube whose length is
much larger than its diameter).

\item The concentration of the outer electrolyte ($a_0$) is assumed 
to be much larger than that of the inner electrolyte ($b_0$) and 
it is also assumed that $a(0,t)$ is kept fixed at $x=0$, [$a(0,t)=a_0$]. 
In experiments one typically has $0.005\le b_0/a_0\le 0.1$.

\item Reaction zones and the precipitation bands are 
much narrower than the diffusion length.

\item  All concentration profiles can be approximated by straight lines
in the neighborhood of the reaction zones and precipitation bands. 

\item The slopes of the straight lines are determined by the 
diffusion lengths and by the asymptotic concentrations of the reacting 
species.

\item An existing precipitation 
band acts as a sink (with $b(x_n,t>t_n)=0$ or $c(x_n,t>t_n)=0$  
boundary condition) for the 
inner electrolyte and for the intermediate 
compound (this is an assumption about the  reaction and  aggregation rates
being sufficiently large).

\end{enumerate}
\vspace{0.2truecm}

Using the above assumptions, we can derive relatively simple
expressions for $p(a_0,b_0)$ for all three classes of theories 
(in Sec.2 for supersturation of ion-product, in Sec.3 for nucleation and 
droplet growth, and in Sec.4 for induced sol-coagulation theories). 
The results are compared both with the corresponding  
numerical solutions of the non-simplified models and with 
experimental data. Sec. 5 contains our final remarks. 
\vspace{0.2truecm}

In closing the introduction, we list the symbols frequently used in the 
text:
\vspace{0.2truecm}

$a(x,t)$ - concentration of outer electrolyte, $A$;

$b(x,t)$ - concentration of inner electrolyte, $B$;

$a_0\, , \, b_0$ - initial concentrations of $A$ and $B$;

$\kappa=b_0/a_0$ - dimensionless ratio of initial concentrations;

$c(x,t)$ - concentration of sol (or intermediate compound);

$D_a, \, D_b, \, D_c$ - diffusion coefficients of $A$, $B$, and $C$;

$D_f$ - effective diffusion coefficient of reaction front ($A+B\to C$);

$q^*$ - ion-product threshold;

$c^*$ - nucleation (coagulation) threshold;

$a^*$ - threshold for induced coagulation; 

$x_n$    - position of the $n$-th precipitation band;

$t_n$  - time of formation of the $n$-th band;

$1+p=x_{n+1}/x_n$ - spacing coefficient.

\section{Ion-product supersaturation}

Let us begin with the simplest case of the reaction $A+B \to D$.
The ion-product supersaturation theory 
\cite{{Ostwald},{Wagner},{Prager},{zeldo}} 
is based on the assumption that 
precipitation of $D$ at time $t$ and at a site $x$ occurs 
if the product of ion concentrations $a(x,t)b(x,t)$
exceeds a threshold value
\be 
a(x,t)b(x,t)\ge q^* \quad .
\label{crition}
\ee
In the mean-field theory, the equations describing this reaction-diffusion 
process are given by
\begin{eqnarray}
\frac{\partial a}{\partial t}&
=& D_a \frac{\partial^2 a}{\partial x^2}  -k\theta(ab-q^*) 
-\lambda   abd ,\label{prager_uno}\\
\frac{\partial b}{\partial t}&
=& D_b \frac{\partial^2 b}{\partial x^2}  -k\theta(ab-q^*)
-\lambda   abd, \label{prager_due}\\ 
\frac{\partial d}{\partial t}&
=&k\theta(ab-q^*)+\lambda   abd,\label{prager_tre}
\end{eqnarray}
where $D_a$ and $D_b$ are the diffusion constants of $A$ and $B$, and 
$\theta(x)$ is the step function describing an infinitely sharp 
threshold for precipitation with $k$ being the rate constant. 
The last terms on the right hand sides represent the 
aggregation  onto the existing precipitate of concentration $d(x,t)$. 
The initial conditions to the above equations are given by 
$a(x,0)=a_0\theta(-x)$, $b(x,0)=b_0\theta(x)$, and $d(x,0)=0$.

These equations can be solved numerically and our analytical 
expression found for $p(a_0,b_0)$ in the above process is
in accord with the numerical results. Before deriving the results, however,
let us show how the structure of the result may emerge from a simple 
dimensional analysis.

If we assume that both the  precipitation ($k$) and aggregation 
($\lambda$) rates go
to infinity then it follows from dimensional considerations that the 
dimensionless 
spacing coefficient can  be expressed through the available 
dimensionless combinations of the parameters and so we have 
\be
p=P(\frac{b_0}{a_0}; \frac{D_b}{D_a}; \frac{q^*}{a_0b_0}) 
\ee
Further simplification can be made by assuming that $P$ is 
not singular in its first argument and setting this argument 
to zero  since in experiments $b_0/a_0 \ll 1$. Then the Matalon-Packter 
law can emerge only if $P$ is linear in its third argument thus giving 
\be
p=P_1(\frac{D_b}{D_a})+P_2(\frac{D_b}{D_a})\frac{q^*}{a_0b_0} \quad .
\ee
We emphasize that this is not a derivation of the  desired result.
This is, however, what we shall show to be valid in the experimentally 
investigated range ($0.05 < p<0.4$) of possible $p$-s. 
Our calculation is based on  a simple analytical approach combined
with numerical evaluation of $P_1$ and $P_2$.

\begin{figure}[htb]
\centerline{
        \epsfxsize=8cm
        \epsfbox{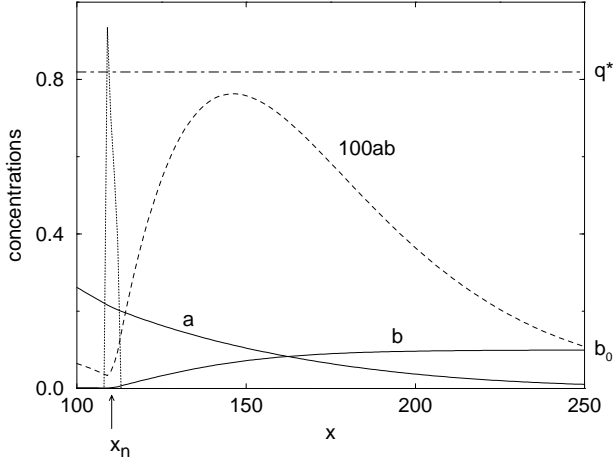}
           }
\vspace{0.5cm}
\caption{Concentration profiles in the ion-product supersaturation model of
Liesegang phenomena. Concentrations are measured in units of $a_0$ while 
the unit of length is $\sqrt{D_a/ka_0}$. The ionproduct, $ab$, and its critical 
value $q^*$ are measured in units of $a_0^2$ and they are magnified by 
a factor 100 for better visibility.}
\label{Fig1}
\end{figure}

Figure~\ref{Fig1} gives a characteristic picture showing the 
concentration profiles as well as the ion-product profile past the 
formation of the precipitation  band $x_n$ and just before 
the appearance of the $({n+1})$-th band.
It is clear from the figure that the linearization of the concentration 
profiles is a good assumption and accordingly, in the neighborhood 
of $x_n$, we can  approximate the concentrations as
\bea
a(x,t)=&a_0\left(1-\frac{x}{\sqrt{2D_at}}\right) \quad ,\\
b(x,t)=&\frac{b_0}{\sqrt{2D_b(t-t_n)}}(x-x_n) \quad .
\eea
The condition for the $n+1$-st band to appear is that the maximum of the
$ab$ product reaches the value $q^*$. This leads to the following two equations
\bea
a(x_{n+1},t_{n+1}) b(x_{n+1},t_{n+1})= q^* \label{prag1} \\
\frac{\partial}{\partial x}[a(x,t)b(x,t)]_{x_{n+1},t_{n+1}}=0\label{prag2}
\eea
The second equation gives a simple relationship between $t_{n+1}$, $x_n$ and 
$x_{n+1}$
\be
\sqrt{2D_at_{n+1}}=2x_{n+1}-x_n
\ee
which allows us to eliminate  $t_{n+1}$ from equation (\ref{prag1}) 
and to obtain an equation for $x_{n+1}/x_n$ and $x_n/x_{n-1}$. 
Assuming now that $n$ is large enough that we can use the  approximation
\be
\frac{x_{n+1}}{x_n}\approx \frac{x_n}{x_{n-1}} =1+p \quad ,
\ee
and one obtains the following equation for $p$:
\be
\Phi (p)=
{p^{3/2}(1+p)\over (1+2p)^2\sqrt{2+p}}
=\sqrt{{D_b\over D_a}}{q^*\over a_0b_0}
\label{pragmatpack1}
\quad .
\ee
This equation clearly has a $p>0$ solution for small values of 
$\zeta \equiv \sqrt{D_b/D_a}q^*/(a_0b_0)$. $p$ increases 
with $\zeta$ and
and  goes to infinity  for  $\zeta\rightarrow 1/4$, accounting for 
the fact that no bands can be formed if $a_0b_0$ is too small for 
$ab$ to reach the critical value $q^*$.

\begin{figure}[htb]
\centerline{
        \epsfxsize=8cm
        \epsfysize=6cm
        \epsfbox{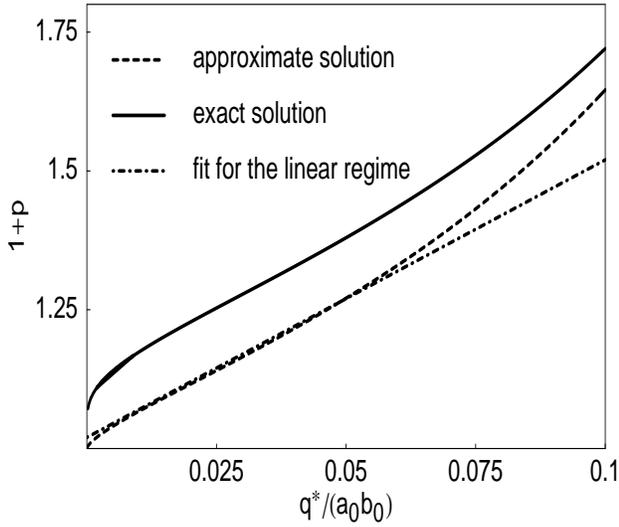}
           }
\vspace{0.3cm}
\caption{Dependence of spacing coefficient $1+p$ on $q^*/(a_0b_0)$ 
as given by the ion-product supersaturation theory for $D_a=D_b=1$. 
The full line corresponds to the solution of 
equations (\ref{prager_uno},\ref{prager_due}), 
the dashed line to our approximate theory. 
The dot-dashed line shows the linear dependence 
for small $q^*/(a_0b_0)$.}
\label{Fig2}
\end{figure}

At first sight, the Matalon-Packter law does not follow from this theory,
since $p$ depends on $a_0$ and $b_0$ only through $a_0b_0$. 
Furthermore, another problem is that, for small $p$, we obtain 
Wagner's result \cite{Wagner}, namely the dependence of $p$ 
on $a_0$ is of the form 
\be
p\sim a_0^{-2/3} 
\ee 
in contrast to the Matalon-Packter prediction $p\sim 1/a_0$. However, 
one should notice that the range of $p$ where $p\sim  a_0^{-2/3}$ holds
is  certainly much smaller than the experimentally 
observed values $0.05<p<0.4$. In this latter interval, $\Phi(p)$ is
very well approximated as a straight line (see Fig.\ref{Fig2})
with parameters $\Phi(p)=-0.0035+0.2p$. Thus, to a good 
approximation, we can write (\ref{pragmatpack1}) as a Matalon-Packter 
law with $F(b_0)=0.02$ and $G(b_0)\sim 1/b_0$
\be
p=0.02 + 5\sqrt{{D_b\over D_a}}{q^*\over a_0b_0} \quad .
\ee
In this approach we obtain $P_1=\rm const.$ and $P_2 \approx 
\sqrt{{D_b}/{D_a}}$.
Numerical solution of the equations~(\ref{prager_uno},\ref{prager_due}) 
indicates that, in reality, the $p$ 
is given to an excellent accuracy by
\be
p \approx 0.18 - 0.052 (\frac{D_b}{D_a})^{1/3}+\left[7.5-2.57 
(\frac{D_b}{D_a})^{1/6}\right]\frac
{q^*}{a_0b_0}
\label{mataionproduct}
\ee

For $D_a=D_b$, the order of magnitude for $F(b_0)\sim 0.12$ agrees
with the order of magnitude values of $F$ in experiments
\cite{Matalon} and the $1/a_0$ dependence is also the 
experimentally observed behavior. However, in experiments,
$F(b_0)$ is not a universal number. 
It changes not only as $D_b/D_a$ is varied but it also depends on $b_0$ 
\cite{Matalon}. Furthermore, the experimentally observed 
$G(b_0)$ is also more varied than being just $G(b_0)\sim 1/b_0$.
Thus we must conclude that although it is possible to derive the 
Matalon-Packter law from the ion-product supersaturation theory, 
the resulting functions $F(b_0)\sim {\rm const}$ and $G(b_0)\sim 1/b_0$ 
are too simple to account for experimental observations.

\section{Nucleation and growth}
\label{nuclandgrowth}

The theory at the next level of complexity 
assumes that the $A+B$ reaction  yields the precipitate $D$ through an
intermediate compound $C$. This compound may, in principle, be a 
molecule $AB$ but it may as well be a sol particle
formed by $A$ and $B$ and, possibly, by some other background ions. 
All this rather complicated situation is described 
by assuming that the $C$-s can be treated as a diffusing species 
which precipitates if its concentration exceeds a threshold $c^*$. 
The precipitate, 
$D$, grows by collecting the neighboring $C$-s and the various 
theories differ in the sophistication of the description of this 
nucleation and growth process \cite{{dee},{luthi}}. 
In our case the nucleation threshold is sharp at $c^*$ and precipitate 
will be assumed to be a perfect sink for the $C$-s.

\begin{figure}[htb]
\centerline{
        \epsfxsize=8cm
        \epsfbox{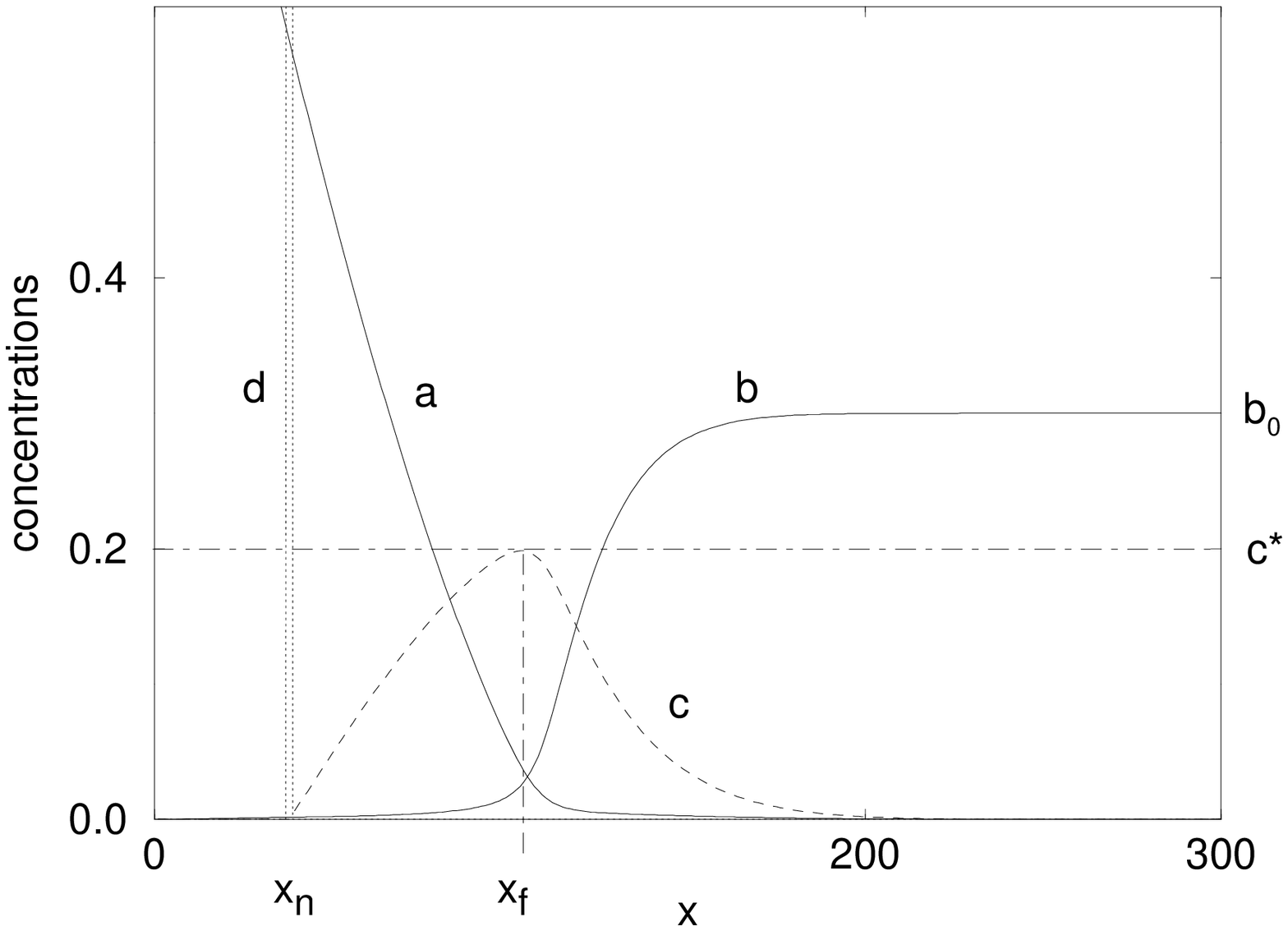}
           }
\vspace{0.3cm}
\caption{Concentration profiles in the nucleation-and-growth 
model of Liesegang phenomena. Units are the same as in Fig.1}
\label{Fig3}
\end{figure}

In order to understand the formation of precipitate, we observe that
the reaction zone (where the $C$-s are produced) is narrow \cite{GR} 
and that $c(x,t)$ reaches its maximum there (see
Fig.{\ref{Fig3}} for characteristic concentration profiles just before
the $n+1$-st band forms). The $n$-th band acts
as a sink  and thus the $C$-s (or about half of them) formed
at the front, $x_f$, end up in this band.
\vspace{0.3cm}
With a good approximation, we can assume that at time $t_{n+1}$,  when $c$  
reaches the threshold value $c^*$ and the $(n+1)$-th 
band is about to form, the concentration of $C$-s exhibits a 
triangular-like shape and it varies linearly between 
$c(x_n)=0$ and $c(x_{n+1})=c^*$. As a consequence, we can estimate the 
current flowing to the $n$-th band as
\be 
\vert j \vert =D_c\frac{c^*}{x_{n+1}-x_{n}} \quad .
\label{current}
\ee
Next we note that
the front becomes quasistationary in the large-time 
limit (its velocity goes as ${\dot x}_f\sim 1/\sqrt{t})$ and so we can 
assume that the above expression for the current is valid 
in the whole time interval 
$[t_n, t_{n+1}]$ (numerical solutions of the reaction-diffusion 
equations support this assumption). Then the amount of $C$ which disappears 
into the $n$-th band during the time interval $[t_n, t_{n+1}]$ can be
computed as 
\be
N_C\approx \vert j \vert (t_{n+1}-t_n)=
D_c c^*\frac{t_{n+1}-t_n}{x_{n+1}-x_{n}}\quad .
\label{N_C}
\ee
On the other hand, $N_C$ can 
also be estimated as the amount of $C$-s produced by the front 
[$c_0(x_{n+1}-x_{n})$, where $c_0$ is a constant \cite{GR}] minus 
the $C$-s which are in the triangle shape, $c^*(x_{n+1}-x_{n})$:
\be
N_C\approx (c_0-c^*)\cdot (x_{n+1}-x_{n}) \quad .
\label{N_C2}
\ee
Equating the two estimates of $N_C$ then yields 
\be
\frac{D_c c^*}{c_0-c^*}=\frac{(x_{n+1}-x_{n})^2}{t_{n+1}-t_n}\quad .
\label{punono2}
\ee
We can use now the fact that the band is formed at the front whose 
position, $x_f$, is determined by an effective diffusion coefficient $D_f$
\cite{GR}
\be
 x_{n+1} \equiv x_f(t_{n+1})=\sqrt{2D_ft_{n+1}}\quad , 
\label{xfront} 
\ee
and find 
\be
\frac{p}{1+p/2}\approx p=\frac{D_c c^*}{D_f(c_0-c^*)}\quad . 
\label{peq}
\ee

\begin{figure}[htb]
\centerline{
        \epsfxsize=8cm
        \epsfysize=6cm
       \epsfbox{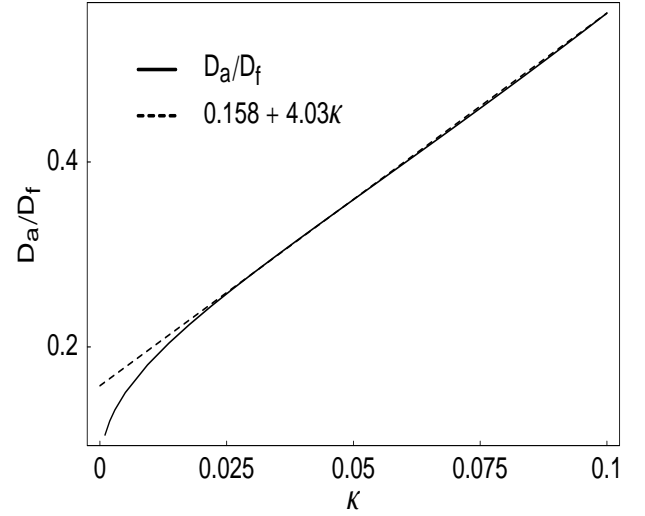}
           }
\vspace{0.3cm}
\caption{Inverse of the effective diffusion constant of the front $D_f$ as a 
function of $\kappa=b_0/a_0$, for $D_a=D_b=1$.}
\label{Fig4}
\end{figure}

The parameters $D_c$ and $c^*$ in the above 
expression are material parameters. The 
dependence of $p$ on $a_0$ and $b_0$ enters through 
the effective diffusion coefficient $D_f$ of the front 
and through the concentration, $c_0$, of $C$-s left 
behind by the moving front.
The remaining task is thus to determine $D_f$ and $c_0$. 
We start with the case of $D_a=D_b$ which can be discussed with relative
ease \cite{GR}. It has been shown for this case that $D_f/D_a$ is 
determined from the following equation:
\be
{\rm erf}\left(\frac{D_f}{2 D_a}\right)=\frac{1-\kappa}{1+\kappa} \quad, 
\label{df}
\ee
where ${\rm erf}(x)$ is the error function \cite{Abram} and $\kappa=b_0/a_0$. 
The remarkable feature of the solution of this
equation is that, for the experimentally 
relevant values $0.005\leq \kappa\leq 0.1$, the inverse of $D_f$ is very well
approximated by a linear function of $\kappa$ (see Fig.\ref{Fig4})
\be
\frac{D_a}{D_f}\approx \eta_1+\eta_2\cdot\kappa
\approx 0.158+4.03\cdot\kappa \quad .
\label{dfinv}
\ee 
This is an important observation since it follows from (\ref{peq}) that
$p\sim 1/D_f$ for small $p$ and the Matalon-Packter law will emerge 
as a consequence of the above equation.

The general case $D_a \not= D_b$ is more complicated but the 
results are similar. The effective diffusion coefficient, $D_f$ 
is obtained from solution of the following equation \cite{koza}
\be
H \Big(-\sqrt{\frac{1}{2}\frac{D_f}{D_a}}\, 
\Big)=\frac{a_0}{b_0}\sqrt{\frac{D_f}{D_a}}
H \Big(\sqrt{\frac{D_a}{2D_b}\frac{D_f}{D_a}}\,  \Big) \label{diffront}
\ee
where $H(x) \equiv [1-erf(x)]\exp(x^2)$. The behavior of $D_a/D_f$ as a 
function of $\kappa$ for  $D_b=D_a=1$ 
is given on Figure~\ref{Fig4}. We see that $D_a/D_f$ is linear 
in $\kappa$ in the experimentally relevant region and thus  
equation (\ref{dfinv}) emerges again but the constants 
$\eta_1$ and $\eta_2$ are now dependent on $D_b/D_a$.

We turn now to the calculation of $c_0$.
The starting point is the observation that the width of the reaction zone 
is much smaller than the diffusion length \cite{GR} and thus  
the reaction zone can be approximated as a point 
at $x_f(t)=\sqrt{2D_ft}$ where both concentrations $a$ and $b$ approach zero. 
This means that the field $b(x,t)$ satisfies 
the diffusion equation with the reaction term replaced by the  
boundary condition $b(x=x_f,t)=0$  and the other boundary 
condition being $b(x\to \infty ,t)=b_0$. 
Once this moving boundary problem is solved, 
the number of $C$-s, $N_C$, produced up to a given time is obtained
by the time integral of the current $|j_b|=D_b\partial b/\partial x$ 
evaluated at $x_f$. Finally, $c_0$ is found by dividing $N_C$ by the advance
of the front in time $t$, $c_0=N_C/x_f$ (here we use the fact that 
the density of $C$-s produced by the front is constant in space~\cite{GR}).
The result of this calculation is given by 
\be
c_0=\frac{b_0}{\sqrt{\pi}} \sqrt{\frac{2D_b}{D_f}}
\exp{\Big(-\frac{D_f}{2D_b}\Big)}
\left[1-{\rm erf}\Big(\sqrt{\frac{D_f}{2D_b}}\,\Big) \right]^{-1}\quad .
\label{c_0exact} 
\ee
For $D_a=D_b$ and $\kappa\ll 1$ we can use equation (\ref{df}) in conjunction 
with the asymptotics of the error function \cite{Abram} 
to obtain $c_0=b_0$.  
The physical meaning of this result is clear. For $\kappa\ll 1$, we have
$D_b\ll D_f$ and thus the front moves fast into the region of $B$-s
and, consequently, the $B$-s can be treated as 
immobile particles yielding $c_0=b_0$. Corrections to the 
$c_0=b_0$ result can also be calculated using again the 
asymptotics of the error function \cite{Abram}. To first order 
in $D_b/D_f$, we find
\be 
c_0\approx b_0\left(1+\frac{D_b}{D_f}\right)=
b_0\left(1+\frac{D_b}{D_a}\frac{D_a}{D_f}\right)  \quad .
\label{c_0asymp}
\ee
We can use now the linearity of $D_a/D_f$ in $\kappa=b_0/a_0$ to 
obtain the following parametrization for $c_0$:
\be
c_0=b_0\left(\sigma_1 + \sigma_2 \kappa\right) \quad ,
\label{c_0param}
\ee
where $\sigma_1 = 1+\eta_1D_b/D_a$ and $\sigma_2=\eta_2D_b/D_a$ 
are numbers of the order of $1$. Since $\sigma_1$ and $\sigma_2$ 
are of the same order of magnitude and since $\kappa\ll 1$, in the 
following we shall neglect the $\kappa$ term in (\ref{c_0param})
\be
c_0\approx \sigma_1 b_0  \quad .
\label{c_0final}
\ee

It is important to note here that a similar omission of 
the $b_0/a_0$ term would not be justified 
in the linearized form of $D_a/D_f$ (\ref{dfinv}).
There we have a constant term, $\eta_1\approx 0.2$ 
that is an order of magnitude smaller than coefficient, $\eta_2\approx 
4$, in front of $b_0/a_0$. Thus, for the relevant values of $b_0/a_0$, 
the two terms contribute equally. 

The expressions (\ref{c_0asymp},\ref{c_0param},\ref{c_0final}) remain good 
approximations up to the point
where $D_b/D_f\approx 1$. This happens, however, only at 
rather large values of $D_b/ D_a$ for $\kappa\ll 1$. Since the diffusion 
coefficients of usual electrolytes in aqueous solutions are usually 
within a factor $2$ of each other \cite{electrolytes1} and their ratios
rarely exceed $5$  \cite{electrolytes2}, 
we believe that the approximation (\ref{c_0param}) can be used  
for realistic experimental setups.
\begin{figure}[htb]
\centerline{
        \epsfxsize=8cm
        \epsfbox{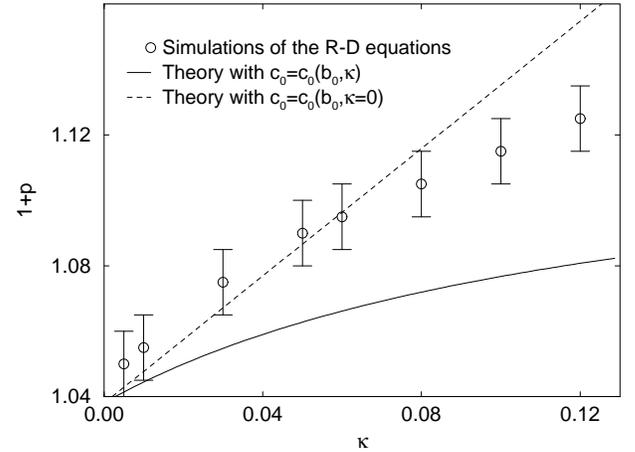}
           }
\vspace{0.3cm}
\caption{Spacing coefficient for the nucleation-and-growth theory. 
The $\kappa=b_0/a_0$ dependence of $p$ is shown for the following choice 
of parameters: $D_b/D_a=1$, $D_c/D_a=0.2$, $c^*/b_0=0.633$ and with all the 
reaction rates taken to be large. The dashed line is the "linear"
Matalon-Packter law (\ref{matanucl}) while the solid line is the 
"non-linear" version of (\ref{matanucl}) where the $\kappa$-dependence 
of $c_0$ (\ref{c_0param}) is kept.} 
\label{Fig5}
\end{figure}
Having determined $D_f$ and $c_0$,  we can now return 
to equation~(\ref{peq}) for $p$ and find again a version of the 
Matalon-Packter law:
\be
p=\frac{D_c c^*\eta_1}{D_a(\sigma_1b_0-c^*)}+
\frac{D_c c^*\eta_2b_0}{D_a(\sigma_1b_0-c^*)a_0} 
=F(b_0)+\frac{G(b_0)}{a_0} . 
\label{matanucl}
\ee
In order to gain confidence in the above result, we have calculated 
$p$ for $\kappa$ in the experimental range $0.005<\kappa<0.1$ by solving
using the appropriate reaction-diffusion equations 
[equations~(\ref{prager_uno},\ref{prager_due},\ref{prager_tre}) 
must be modified
and supplemented by another equation in order to take into account
the nucleation and growth processes described at the start of the section
(for details see \cite{luthi})].
The results are displayed on Figure~\ref{Fig5} where the
Matalon-Packter law derived above (solid line on the figure)
is shown to perform very well 
considering the simplicity of the derivation.  It should be noted that 
nonlinear dependence on $\kappa$ sets in for $\kappa\ge 0.06$ and that
taking into account the $\kappa$-dependence of $c_0$ (dashed line) 
modifies the straight line with the right curvature. 

One can observe from equation (\ref{matanucl}) that 
the nucleation-and-growth theory gives more complicated
$b_0$ dependences for $F(b_0)$ and $G(b_0)$ than the 
ion-production theory. Power law forms for $F$ and $G$
are found in the limit of $c^*\ll b_0$ (which might be the experimental
limit) where one obtains $F(b_0)\sim 1/b_0$ and $G(b_0)\sim \rm constant$. 
It is remarkable that these are similar to the ion-product results but with
the roles of $F$ and $G$ interchanged.
Comparing the two theories, we see that the nucleation-and-growth theory 
performs better in the sense that it can produce 
power law behavior for $F(b_0)$ as observed in some experiments and, 
at the same time, has a $G(b_0)$ which depends on $b_0$ more weakly 
but it is a decreasing function of $b_0$ in agreement with experiments.

In conclusion,  the nucleation-and-growth theory provides us with 
a Matalon-Packter law that is closer to experiments and, perhaps, describes
some of them. Ne\-vertheless, the functions $F$ and $G$ appear  
to be too rigid to accomodate all experimental findings. It should also 
be noted that this theory has a prediction $G(b_0)/F(b_0)\sim b_0$ 
that is experimentally easily distinguishable  
from the prediction of the ion-product theory, $G(b_0)/F(b_0)\sim 1/b_0$.

\section{Induced sol-coagulation}
\label{indsolcoag}

The {\em induced sol-coagulation} theory \cite{{Chatter},{shino}} 
is a generalization of the {\em nucleation-and-growth} theory. 
The sol, $C$, is produced at the reaction front in the $A+B\to C$ 
reaction and it flocculates if the following two conditions are satisfied.
First, $c$ must exceed a supersaturation 
threshold $c^*$ and, second, the concentration of the outer electrolyte, 
$a$, must be above the critical coagulation concentration threshold, $a^*$.
The second condition arises in systems where  
the $A$ ions screen 
the repulsive electrostatic interaction among the sol particles. 

\begin{figure}[htb]
\centerline{
        \epsfxsize=8cm
        \epsfbox{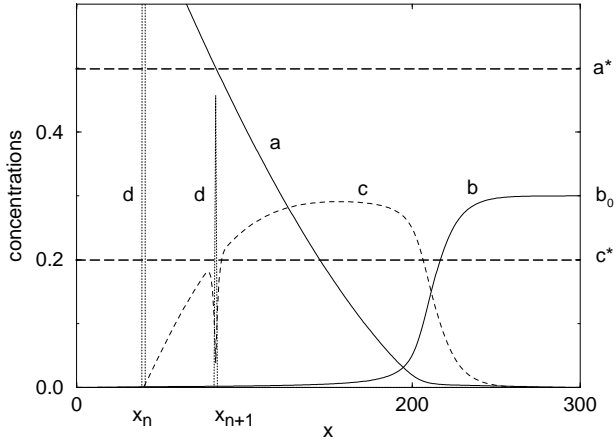}
           }
\vspace{0.3cm}
\caption{Concentration profiles and thresholds in the 
induced sol-coagulation theory. Units are the same as in Fig.1.}
\label{Fig6}
\end{figure}

Fig.\ref{Fig6} shows characteristic concentration profiles in this 
process just as
the $n+1$-st band is appearing. The first remarkable feature of this 
picture is that the reaction front is way ahead of the precipitation
zone (this does seem to happen in some experiments \cite{kaimul}). 
This is understandable, since $a(x_f,t)\rightarrow 0$ 
at the front while at the place where  precipitation
occurs one must have $a(x_{n+1},t_{n+1})\ge a^*$. The second important 
feature is that the reaction front leaves $C$-s behind 
at a fixed concentration $c_0$, as already discussed 
in the previous section. In order that precipitation could occur,
the parameters must be assumed such that $c_0>c^*$. 
Further assuming that the sol does not diffuse too
fast $(D_c\ll D_a)$, one of the condition 
for precipitation ($c\ge c^*$) is always satisfied behind the front and
far away from the last existing band $(x_n)$.

Consequently, the position, $x_{n+1}$, 
where the next band appears will be determined by the arrival of the 
concentration `front' $a=a^*$ to a position where $c=c^*$. Since the $a$ and 
$c$ profiles near $x_n$ can be written as 
$a=a_0(1-x/x_f)$ and $c=c_0(x-x_n)/\sqrt{2D_c(t-t_n)}$,
the above conditions yield
\bea
{{c_0(x_{n+1}-x_n)}\over {\sqrt{2D_c(t_{n+1}-t_n)}}}=&c^* \quad ,
\label{fleq1}\\
a_0\left(1-{{x_{n+1}}\over {x_f(t_{n+1})}}\right)=&a^*  \quad ,
\label{fleq2}
\eea
where $x_f(t)=\sqrt{2D_ft}$ is the position of the reaction front at time $t$.  

In order to calculate $p$, we note now that equation (\ref{fleq2}) yields
\be
\frac{x_{n+1}}{x_f(t_{n+1})}=1-\frac{a^*}{a_0}=\frac{x_n}{x_f(t_{n})} 
\quad .
\label{fleq3}
\ee
The above equation can then be used in conjunction with  
$x_f(t_{n+1})=\sqrt{2D_ft_n}$ to obtain $p$ from equation (\ref{fleq1})
\be
\frac{p}{1+p/2}\approx p=\frac{2D_c{c^*}^2}{D_fc_0^2(1-a^*/a_0)^2} \quad .
\label{fleq4}
\ee
Since the values of $D_f$ and $c_0$ come from reaction $A+B\to C$ which is 
decoupled from the the next stages of nucleation and growth processes,
the calculations and approximations of these 
quantities can be taken over from Section \ref{nuclandgrowth}.
Moreover, we shall assume that $a^*$ is much smaller than $a_0$ 
so that the right-hand 
side of eq.(\ref{fleq4}) can be expanded in $a^*/a_0$. There is  
no experimental information on $a^*$ and, although $a^*/a_0\ll 1$
seems to be a natural assumption, we must be 
cautious with the $a^*/a_0\to 0$ limit. 
The problem is that  the picture we are working with for the 
induced sol-coagulation theory (Figure \ref{Fig6}) 
is not valid for arbitrary small $a^*$. Indeed, the fact that the position
of the $A+B\to C$ reaction front, $x_f$ is far ahead 
of the point of formation of
the $n+1$-st band means that diffusion length of the $C$ particles 
is smaller than $x_f(t_{n+1})-x_{n+1}$
\be
\sqrt{2D_c(x_{n+1}-x_n)}<x_f(t_{n+1})-x_{n+1} \quad .
\label{a*/a0condition}
\ee
Using equations (\ref{fleq3},\ref{fleq4}) and 
neglecting higher order terms in $p$ and $a^*/a_0$, 
the above equation yields 
\be
\frac{D_cc^*a_0}{D_fc_0a^*}<\frac{1}{2} \quad .
\label{a*limit}
\ee
The meaning of this result is the following. The place where the 
new band nucleates moves to the $A+B\to C$ reaction zone and thus 
the inequality (\ref{a*/a0condition}) gets violated upon 
increasing $D_c$ or $c_*$ and decreasing $a^*$. 
It is clear thus that, at fixed values of the 
other parameters,  $a^*/a_0$ cannot be
taken to zero. 
\begin{figure}[htb]
\centerline{
        \epsfxsize=8cm
        \epsfbox{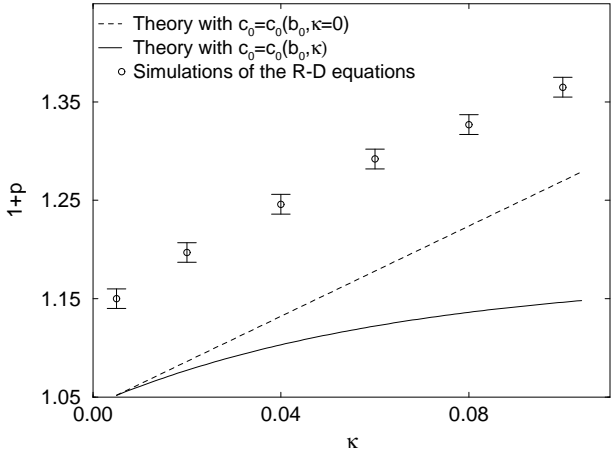}
           }
\vspace{0.3cm}
\caption{Spacing coefficient for the induced sol-coagulation theory. 
The dependence of $p$ on $\kappa=b_0/a_0$ is displayed for the following 
choice of parameters: $D_b/D_a=1$, $D_c/D_a=0.3$, $c^*/b_0=0.86$,  
$a^*/b_0=7.8$, and with all the 
reaction rates taken to be large. The dashed line is the "linear"
Matalon-Packter law (\ref{matainduced}) while the solid line is the 
"non-linear" version of (\ref{matainduced}) where the $\kappa$-dependence 
of $c_0$ (\ref{c_0param}) is kept when going from eq.(\ref{fleq4}) 
to eq.(\ref{matainduced}).}
\label{Fig7}
\end{figure}
The lack of the limit $a^*/a_0\to 0$ has the consequence that 
the result for the 
nucleation-and-growth theory (\ref{matanucl}) cannot be obtained from 
(\ref{fleq4}). The matter of finding 
an expression for $p$ valid in the $a^*/a_0\to 0$ limit and
reducing to (\ref{matanucl}) 
appears to be a rather difficult task since various lengthscales become
comparable approaching this limit. 
We have not been able to obtain such an interpolating formula.  

Returning to (\ref{a*limit}), we see that the 
material parameters can be such (e.g. if $c^*/c_0\ll 1$) 
that $a^*/a_0$ can be small without the inequality (\ref{a*limit}) 
being violated. For such range 
of parameters we can expand (\ref{fleq4}) in  $a^*/a_0$ and  
obtain again a version of the Matalon-Packter law:
\be
p=\frac{2 D_c{c^*}^2}{D_a\sigma_1^2}\frac{\eta_1}{b_0^2}+\frac{2 
D_c{c^*}^2}{D_a\sigma_1^2} \cdot
\frac{(2\eta_1a^*+\eta_2 b_0)}{b_0^2a_0}
\label{matainduced}
\ee

The spacing coefficient (\ref{matainduced}) can again be compared with 
that obtained from the appropriate 
reaction diffusion equations (the equations for the nucleation-and-growth
theory must be augmented with the condition that nucleation can occur 
only if $a>a^*$). An exhaustive numerical study is practically impossible 
due to the number of parameters in the problem  
($\kappa$, $a^*/a_0$, $c^*/a_0$, $D_b/D_a$, $D_c/D_a$), plus the rates 
of reaction, nucleation, and aggregation).
One can easily miss regimes of nontrivial behavior 
in this high dimensional parameter space and we claim 
with this numerical study only that the Matalon-Packter law
as given by equation (\ref{matainduced}) is indeed observed 
for a reasonable range of parameters (Figures \ref{Fig7}).
 
As one can see from (\ref{matainduced}), the induced sol-coagulation
theory provides us with 
\be
F(b_0)\sim 1/b_0^2\quad , \quad
G(b_0)\sim (\alpha +\beta b_0)/b_0^2 \quad .
\label{FGsolcoag}
\ee

The power law form $F(b_0)\sim b_0^{-2}$ is close to what has been 
observed in some experiments \cite{Matalon} 
and the fact that $G(b_0)$ is a decreasing 
function of $b_0$ is also in accord with the observations. 
It should also be clear that a more precise theory 
of the  induced sol-coagulation process should 
reproduce, in the $a^*/a_0\to 0$ limit, the result $F\sim b_0^{-1}$ 
obtained in the nucleation-and-growth theory. Thus, assuming that the 
crossover between the $b_0^{-2}$ and the $b_0^{-1}$ behaviours is 
smooth, one should be able to find regimes where $F(b_0)\sim b_0^\gamma$ 
with $1\le \gamma \le 2$.  Since this covers a large
portion of the experimentally observed 
range $0.2\le \gamma \le 2.7$, we conclude that the 
induced sol-coagulation theory provides the best description of the 
pattern formation in Liesegang phenomena.

\section{Final remarks}

Our main results are summarized in equations (\ref{mataionproduct}), 
(\ref{matanucl}), and (\ref{matainduced}) giving the Matalon-Packter 
law for the three main theories, respectively. Comparing these 
formulas, we arrived at our main conclusion, namely 
that the induced sol-coagulation theory is the 
best in describing the experimental observations on the 
spacing coefficient of Liesegang patterns.

It is no doubt that the arguments used in the derivation 
of $p$ can be refined and made more precise. 
The main aim of our work, however, was the demonstration that 
the Matalon-Packter law can be understood in terms of simple 
pictures for which  it is possible to develop analytical
arguments. We hope that one can build on these results and 
achieve a better understanding of Liesegang phenomena.
In particular, it would be important to find a 
description of the induced sol-coagulation process which contained the 
crossover to results for the nucleation-and-growth model.
Furthermore, we have just considered 
the simplest cases associated with the reaction $A+B\to C$. 
In experiments, equally important are 
the  $A_2+B\to C$ and $A+B_2\to C$ cases and, obviously,   
one should also explore the possibilites 
associated with these and, possibly, more complicated reaction schemes. 
Finally, an interesting and important test of the conclusion that the 
induced sol-coagulation theory is preferable in the studies
of Liesegang phenomena would be the application 
of this theory for the {\em quantitative } description of 
revert patterns \cite{{gosh},{kanniah},{flicker},{hantz}}.

\section*{Acknowledgments}
We thank B. Chopard, P. Hantz, and T. Unger for useful discussions.
This work has been partially supported by the 
Swiss National Science Foundation 
in the framework of the Cooperation in Science and Research with CEEC/NIS,
by the Hungarian Academy of Sciences (Grant OTKA T 019451), and 
by the EPSRC, United Kingdom (Grant No. GR/L58088).

\end{multicols}


\begin{references}
\bibitem{revertnote} There are systems that 
display revert patterns, namely the distance between the bands decreases 
with band number instead of increasing as in the normal Liesegang phenomena.
In these cases $x_n/x_{n-1}\to 1$ i.e. $p=0$ and the 
Matalon-Packter law is clearly not applicable.
\bibitem{postnuc} There is a class of theories 
                  \cite{{flicker},{kaimul},{ortoleva},{venzl}} 
                  which attribute the Liesegang band formation to 
                  the instability present
                  in the postnucleation coarsening process. These 
                  theories however cannot explain the regular aspects
                  of the band spacing.
\end{references}
\end{document}